\documentstyle[twoside,ijmpe1,epsfig]{article}
\catcode`\@=11
\long\def\@makefntext#1{
\protect\noindent \hbox to 3.2pt {\hskip-.9pt
$^{{\eightrm\@thefnmark}}$\hfil}#1\hfill}

\def\@makefnmark{\hbox to 0pt{$^{\@thefnmark}$\hss}}

\def\ps@myheadings{\let\@mkboth\@gobbletwo
\def\@oddhead{\hbox{}
\rightmark\hfil\eightrm\thepage}
\def\@oddfoot{}\def\@evenhead{\eightrm\thepage\hfil
\leftmark\hbox{}}\def\@evenfoot{}
\def\sectionmark##1{}\def\subsectionmark##1{}}

\def\qed{\hbox{${\vcenter{\vbox{                        
  \hrule height 0.4pt\hbox{\vrule width 0.4pt height 6pt
  \kern5pt\vrule width 0.4pt}\hrule height 0.4pt}}}$}}

\def\bsc{{\sc a\kern-6.4pt\sc a\kern-6.4pt\sc a}}       
\def\bflatex{\bf L\kern-.30em\raise.3ex\hbox{\bsc}\kern-.14em
T\kern-.1667em\lower.7ex\hbox{E}\kern-.125em X}
\newcommand{\ber}{\begin{eqnarray}}
\newcommand{\eer}{\end{eqnarray}}
\newcommand{\bea}{\begin{equation}}
\newcommand{\eea}{\end{equation}}
\def\be{\begin{eqnarray}}
\def\ee{\end{eqnarray}}
\def\bd{\begin{displaymath}}
\def\ed{\end{displaymath}}

\begin{document}
\centerline{\bf 
Deformation Projected RMF Calculation for Cr and Fe nuclei}
\vspace{0.035truein}
\centerline{\bf
in Hybrid Derivative Coupling Model}
\vspace{0.37truein}
\centerline {\footnotesize PARNA MITRA and
G. GANGOPADHYAY}
\vspace{0.015truein}
\centerline {\footnotesize\it Dept. of Physics, University College of
Science, University of Calcutta}
\baselineskip=10pt
\centerline {\footnotesize\it 92 Acharya Prafulla Chandra Road, Kolkata
700 009,
India}
\date{}
\vspace*{0.21truein}

\abstracts{
The ground state properties of even mass Cr and Fe isotopes are studied using 
the generalized hybrid derivative coupling model. The energy surface of each 
isotope is plotted as a function of the mass quadrupole moment. The neutron 
numbers N=20 and N=40 are seen to remain magic numbers but N= 28 and 50 
are predicted 
to be non-magic. The neutron number N=70 turns out to be a magic number 
according to the present calculation. In all the isotopes studied the 
calculated  binding energy values are less than those obtained from experiment 
while the deformation is in better agreement.
}{}{}
\vskip 4cm

PACS Number(s): 21.60.Jz,27.40.+z,27.50.+e,27.60.+j
\vskip 1cm
Running Head: RMF Calculation for Cr and Fe nuclei
$\ldots$
\newpage
\baselineskip 18 pt

\section{Introduction}
\vspace*{-0.5pt}
\noindent

Recent advances in producing light nuclei close to the drip line using 
radioactive ion beams has produced a renewed interest in the study of these
nuclei. These nuclei, with extreme values of isospin, provide a testing ground 
of different theoretical models. 

Relativistic mean field (RMF) calculations achieved 
considerable success in the description of various ground state 
properties of nuclei including the binding energy, deformation 
and charge radius throughout the periodic table.\cite{1} In a related approach, 
Zimanyi and Moszkowsky used derivative coupling between the scalar meson 
and the nucleon.\cite{2}  One important problem of their model is that it 
predicts a high value of 
the effective nucleon mass and consequently a lower spin orbit splitting. 
A better description of various properties of nuclear matter at zero and 
finite temperature as well as finite nuclei can be obtained using the generalized 
hybrid derivative coupling model\cite{3,4} where both Yukawa point coupling and 
derivative coupling between the baryons and the scalar meson 
were used with a strength ratio $\alpha/(1- \alpha)$ where
$\alpha$ = 0.25.
In a previous work\cite{5} we employed the generalized hybrid derivative coupling 
model to study the ground state of even-even Ne, Mg, Si and S isotopes. The 
deformation values obtained were close to the experimental measurements in most of 
the nuclei studied. In contrast, the conventional RMF models using the NL3 
or NL-SH sets of parameters often fails to accurately predict the deformation
in lighter nuclei.
The success of the model in describing the shape of lighter nuclei 
prompted us to extend our study to Cr and Fe isotopes. The principal aim of 
this work is to study the deformation in these nuclei. 

\section{Calculation and Results}

The details of the method of calculation have 
been presented in elsewhere\cite{4,5} and are not described here.  The starting 
point is a scaled Lagrangian with both scalar and derivative 
$\sigma-N$ couplings. Under the usual
approximations, {\em eg.} classical meson fields,
time-reversal symmetry, point-like nucleons, no sea approximation,
the surviving components for the
the time independent equations, for the meson
fields and the nuclear wave function are given
in our earlier publications\cite{4} and are not repeated here.

We have employed the BCS formalism in a constant gap approach.
The gap parameters are either taken from the odd-even 
mass difference or calculated using the formula 
$\Delta_{n(p)}=11.2/\sqrt{N(Z)}$. 
In the  present work we have assumed axial symmetry 
and plotted the energy surface as a function of the mass quadruple moment. The 
energy surface provides a better understanding of the deformed system.  It 
indicates the possibility of shape co-existence and also, considering the 
shallowness of the minima, the possibility for nuclear shape to be triaxial.  
We have taken the numbers of oscillator shells for both Fermions and Bosons to 
be 12. 
Increasing the number of shells has  very little
effect on the total binding energy or the quadrupole deformation,
even for very neutron rich nuclei.
For example, in $^{82}$Cr, if the number of each type of shell is taken to 
be 14, the total binding energy increases by only 0.3 MeV while there is no 
change in the shape of the energy surface.

In most of the nuclei we have studied, we  have obtained more than one minimum, 
generally corresponding to prolate and oblate shapes. If one of the 
minima is much 
deeper than the other, we have assumed it to be the ground state. If the energies of 
the two are very close to each other and both of them are deep, 
one of them is possibly the true minimum. However, if both 
the minima are very shallow, the nucleus is more likely to be
triaxial and the minima obtained are actually saddle points.

We have calculated the ground state binding energy and deformation values for $^{44-92}$Cr 
and $^{48-96}$Fe, nuclei which are stable against nucleon drip according to the present 
model. Figures 1 and 2 represent the excitation energy surfaces 
as a function of the mass quadrupole deformation for Cr and Fe isotopes, respectively. 
The nature of the curves are similar to those obtained for lighter 
isotopes such as Ne, Mg, Si and S.\cite{5}
 
Table 1 compares the calculated binding energy and deformation for Cr isotopes
with experiment. In the case of Cr isotopes with neutron magic number N=20 and 
N=40, i.e. for $^{44}$Cr and $^{64}$Cr respectively, the ground state solutions
are found to be spherical. In $^{46}$Cr both the minima are very shallow 
indicating possible triaxiality. The ground state of the isotopes $^{48-58}$Cr 
are prolate although the oblate minima becomes progressively deeper with neutron 
number. The next two even Cr isotopes $^{60,62}$Cr are found to be oblate though 
they actually  may be triaxial. 
The ground state of $^{66}$Cr turns out to be spherical. The earlier pattern of 
prolate minima being the ground state while the oblate minima getting deeper 
continues throughout $^{68-84}$Cr. The present work predicts the ground state 
of $^{86}$Cr to be oblate. The isotopes
$^{88,90}$Cr are probably triaxial. The dripline nucleus $^{92}$Cr is nearly 
spherical. In our calculation quadrupole deformation for $^{48-54}$Cr isotopes are 
closer to the experimental values in comparison to the an earlier work.
For example Lalazissis {\em et al}\cite{8} found $^{52}$Cr to be spherical.
In fact, all the available mean field calculations predict this nucleus to
be spherical. In contrast 
our calculation predicts the deformation to be 0.345, closer to the experimentally 
measured value 0.224.  The prediction of the ground state deformation of $^{52}$Cr
can be traced to the low spin-orbit splitting in the present model. In the conventional
RMF calculation at zero deformation, using the parameter set NL3, 
the spin-orbit splitting between the neutron $1f_{7/2}$ and the $1f_{5/2}$ levels comes out 
to be nearly 8 MeV, a value double than that obtained in the present model, i.e. 4 MeV.
Consequently, the shell gap is 4 MeV in the present approach, while conventional
RMF calculations predict it to be nearly 8 MeV. The energy level spacing is 
very similar to that of the other N=28 nucleus studied in the present work,
{\em viz.} $^{54}$Fe, whose single particle energy levels at zero deformation
are  given later in Figure 3.
 The neutron number N=28 can no 
longer  be considered as a magic number. The absence of the shell gap at N=28
signifies that the nucleus $^{52}$Cr lies near neutron midshell and hence show a 
large deformation. All the other neighbouring Cr and Fe isotopes show this feature.

Table 2 compares the calculated and the experimental binding energy and deformation 
for Fe isotopes. In $^{48}$Fe both the prolate and the oblate minima are very shallow. 
Hence the nucleus is likely to have a triaxial shape. In isotopes $^{50-60}$Fe the 
prolate minimum is deeper than the oblate minimum. But the oblate minimum grows 
deeper as the neutron number increases and becomes the ground state at $^{62}$Fe.
The minima are comparatively shallow and the shape may actually be triaxial.
The nucleus $^{64}$Fe 
may also possibly be triaxial in its ground state. The neutron magic nucleus 
$^{66}$Fe with N=40 is found to be spherical.  The ground state of $^{68}$Fe is
nearly spherical. Again from $^{70}$Fe to $^{88}$Fe 
the prolate minimum is deeper indicating it to be the ground state. In $^{90}$Fe 
two minima are nearly equal. In $^{92}$Fe the oblate minimum is slightly deeper 
but both the minima are very shallow indicating possible triaxiality. The nuclei 
$^{94,96}$Fe are spherical. For $^{54}$Fe and for $^{56}$Fe, the deformation is slightly 
greater than experimental value. But for $^{58}$Fe and $^{60}$Fe deformations values 
are predicted with greater accuracy.   

In all the isotopes studied the calculated values for binding energy are less than 
the experimental ones.  This has been observed in our earlier work also.\cite{5} From 
the figures it is seen that both the Fe and Cr isotopes with neutron magic number 
N=28 and 50 
are not spherical. Thus one feature that stands out is that N=28 and 50 are not magic 
numbers in our calculation.  This is due to the fact that the splitting between 
the neutron 1$f_{7/2}$ and 1$f_{5/2}$ levels in the former and between the neutron 
1$g_{9/2}$ and the 1$g_{7/2}$ 
levels in the latter are small. This is shown in figure 3 for $^{54}$Fe and $^{76}$Fe.  
Hence although N=40 is
magic, N=28 and 50 are not. The N=70 nucleus, $^{96}$Fe is spherical in its ground 
state. An examination of the single particle energy levels  given in figure 3 
reveal that  the splitting 
between the neutron $1i_{11/2}$ and $1i_{9/2}$ is small. Thus the intruder orbital $1i_{11/2}$
lies higher up in energy and hence N=70 turns out to be magic.

\section{Summary}
\vspace*{-0.5pt}
\noindent

To summarize we study ground state properties of Cr and Fe isotopes using the generalized 
hybrid derivative coupling model. The energy surface for each isotope is plotted as a 
function of the mass quadrupole moment. The neutron numbers N=20 and N=40 are magic 
numbers but N=50 turns out to be non-magic. The nucleus $^{96}$Fe is predicted to be 
spherical in its ground state and N=70 turns out to be a magic number according to 
our calculation. In all the isotopes studied the calculated  binding energy values 
are less than those obtained from experiment. However, the deformation is better in 
comparison to those in other works.

The authors gratefully acknowledge the grant of a minor research project under University Grants Commission, New Delhi.
\newpage
\nonumsection{References}

\clearpage

\begin{table}[t]
\begin{center}
\caption{
The calculated binding energy and deformation values for Cr isotopes. The results 
of the present calculation are compared with 
experimental values. The experimental binding energy and deformation values are 
from Ref. 6 and Ref. 7, respectively.}
\begin{tabular}{|c|c|c|r|l||c|c|r|}\hline
A &\multicolumn{2}{c|}{B.E. (MeV)}&
\multicolumn{2}{c||}{$\beta$}&
A &\multicolumn{1}{c|}{B.E. (MeV)}&
\multicolumn{1}{c|}{$\beta$}\\
\cline{2-8}
&
Theo.&
Expt.&
Theo.&
Expt.
&&
Theo.&
Theo.
\\\hline
44&7.92&7.95&0.005&&70&7.90&0.259\\
46&8.17&8.30&0.258&&72&7.75&0.341\\
48&8.43&8.57&0.308&0.336&74&7.59&0.300\\
50&8.50&8.70&0.355&0.293&76&7.42&0.286\\
52&8.56&8.78&0.345&0.224&78&7.25&0.319\\
54&8.57&8.78&0.296&0.250&80&7.09&0.265\\
56&8.55&8.72&0.253&&82&6.92&0.240\\
58&8.52&8.64&0.214&&84&6.76&0.217\\
60&8.47&8.54&-0.132&&86&6.62&-0.158\\
62&8.43&8.43&-0.086&&88&6.48&-0.150\\
64&8.36&&0.004&&90&6.34&-0.099\\
66&8.21&&0.005&&92&6.22&0.003\\
68&8.05&&0.223&&&&\\
\hline
\end{tabular}
\end{center}
\end{table}

\clearpage

\begin{table}[t]
\begin{center}
\caption{
The binding energy and deformation values for Fe isotopes. See caption of table 1 for details.} 
\begin{tabular}{|c|c|c|r|l||c|c|r|}\hline
A &\multicolumn{2}{c|}{B.E. (MeV)}&
\multicolumn{2}{c||}{$\beta$}&
A &\multicolumn{1}{c|}{B.E. (MeV)}&
\multicolumn{1}{c|}{$\beta$}\\
 \cline{2-8}
&
Theo.&
Expt.&
Theo.&
Expt.
&&
Theo.&
Theo.
\\\hline
48&7.87&8.03&0.245&&74&8.04&0.334\\
50&8.16&8.35&0.357&&76&7.90&0.342\\
52&8.41&8.61&0.408&&78&7.76&0.350\\ 
54&8.55&8.74&0.411&0.195&80&7.61&0.336\\
56&8.55&8.79&0.349&0.240&82&7.46&0.326\\
58&8.59&8.79&0.305&0.25&84&7.30&0.285\\
60&8.60&8.76&0.265&0.222&86&7.15&0.255\\
62&8.57&8.69&-0.153&&88&7.01&0.213\\
64&8.54&8.61&-0.107&&90&6.86&0.178\\
66&8.51&8.53&0.004&&92&6.74&-0.105\\
68&8.39&&0.004&&94&6.61&0.001\\
70&8.27&&0.218&&96&6.50&0.000\\
72&8.16&&0.277&&&&\\

\hline
\end{tabular}
\end{center}
\end{table}

\clearpage
\centerline{\bf List of figure captions}
\begin{figure}[h]
\vskip 2.0cm
\includegraphics{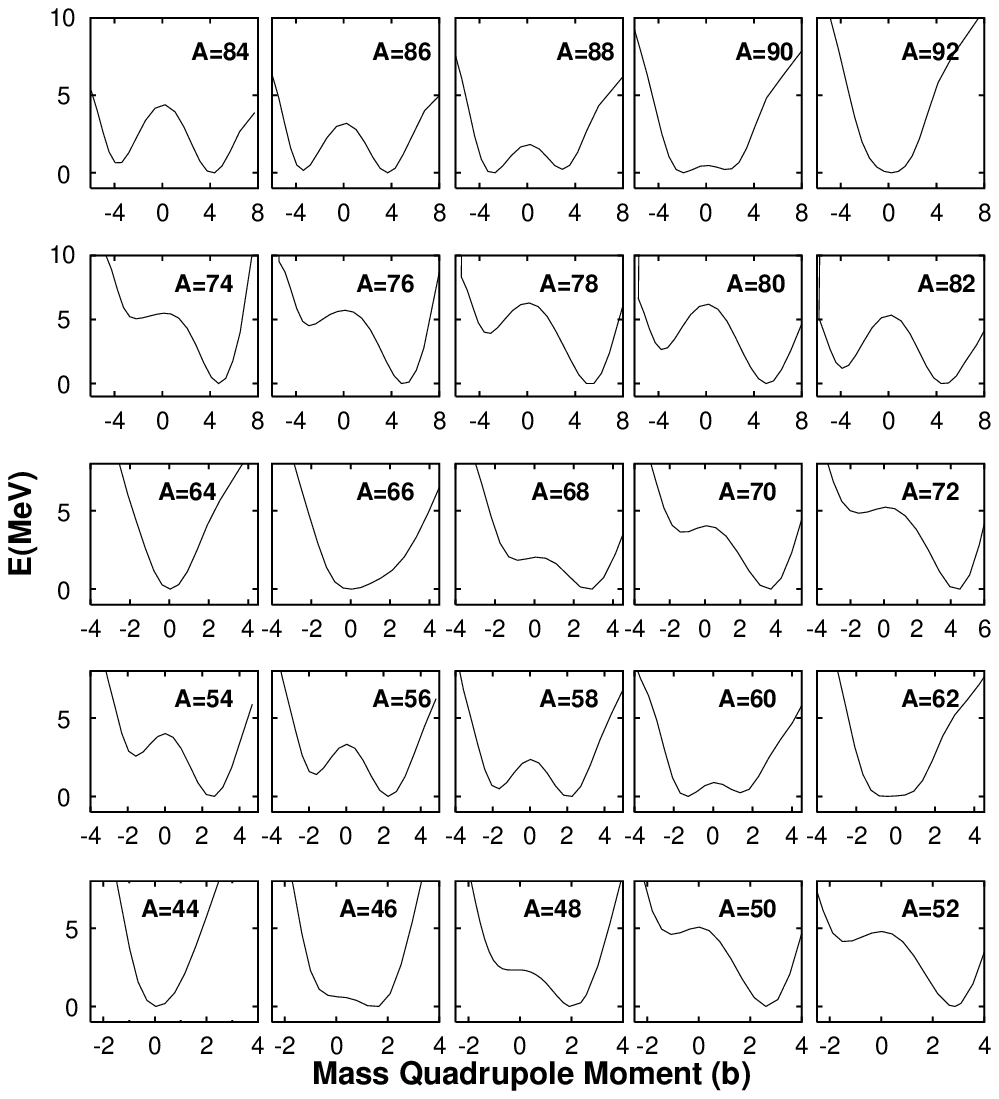}
\caption{\label{fig1}
The excitation energy surfaces of different Cr isotope as a function of the mass quadrupole moment.
}
\end{figure}

\begin{figure}[h]
 \includegraphics{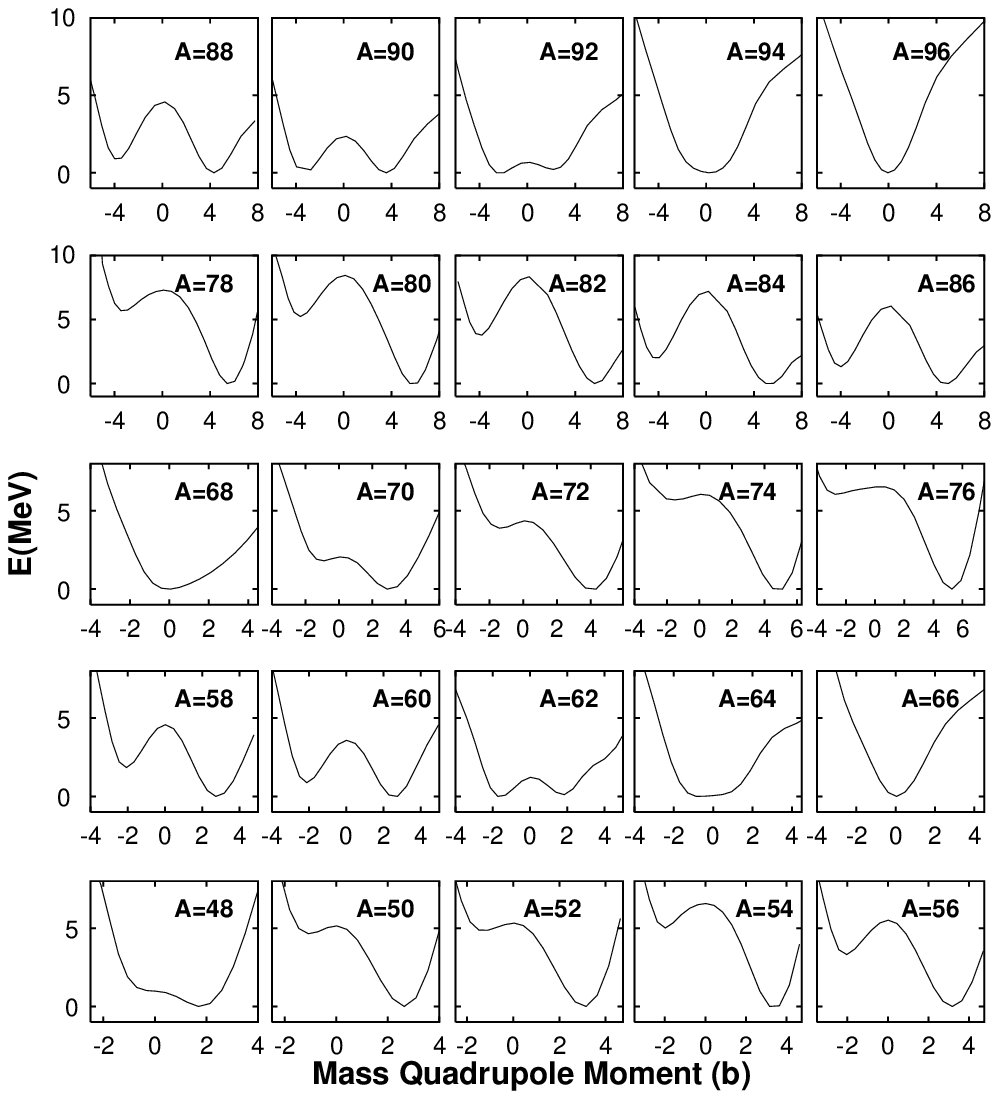}
\caption{\label{fig2}
The excitation energy surfaces of different Fe isotope as a function of the mass quadrupole moment.
}
\end{figure}
\begin{figure}[h]
\vskip -5cm
\parindent -2cm 
\includegraphics{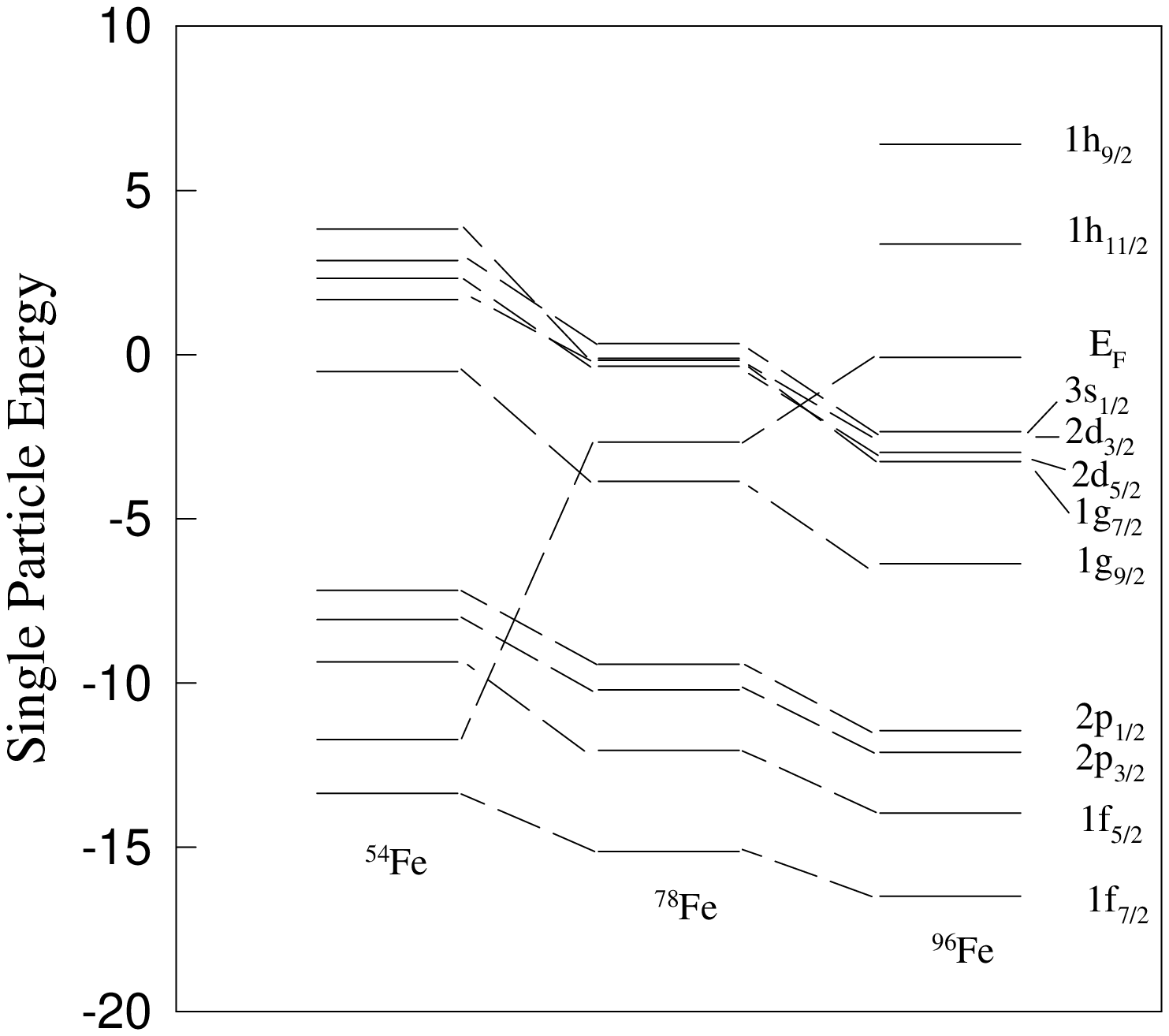}
\vskip -6cm
\caption{\label{fig3}
The neutron single particle energy levels of Fe isotopes at zero deformation 
for N=28, 50 and 70. Here, E$_F$ refers to the Fermi level.}
\end{figure}

\noindent


\begin{thebibliography}{}
\bibitem{1}See {\em eg.} P. Ring, Prog. Part. Nucl. Phys., {\bf 37}, 13 (1996).
\bibitem{2} J. Zimanyi and S. A. Moszkowski, Phys. Rev. C {\bf 42}, 1416 (1990).

\bibitem{3} S.Sarkar and B. Malakar, Phys. Rev. C {\bf50}, 757 (1994)

\bibitem{4} B. Malakar and G. Gangopadhyay, Euro. Phys. Jour. {\bf A10}, 267 (2001); 
P. Mitra, B. Malakar, and G. Gangopadhyay, Int. Jour. Mod. Phys.E {\bf 10}, 475 (2001).

\bibitem{5}P.Mitra, G. Gangopadhyay, and B. Malakar, Phys. Rev. C {\bf 65}, 034329 (2002).

\bibitem{6} G.Audi and A.H. Wapstra, Nucl. Phys. {\bf A595}, 409(1995)
\bibitem{7} S. Raman, C.H. Malarkey, W.T. Milner, C.W. Nestor , and P.H. Stelson, At. Data Nucl. Data Tables {\bf 36}, 1 (1987).

\bibitem{8} G. A. Lalazissis, A. R. Farhan, and M. M. Sharma, Nucl. Phys. A628, 221 (1998)

\end{thebibliography}
\end{document}